%% file: ACO_Paper.tex
\documentclass[letterpaper,twocolumn,10pt]{article}

\usepackage{tikz}
\usepackage{amsmath}
\usepackage{authblk}
\usepackage{blindtext}

\usepackage{hyperref}

\begin{document}

\date{}
\title{\Large \bf CybORG: An Autonomous Cyber Operations Research Gym}

\author[1]{Callum Baillie}
\author[2]{Maxwell Standen}
\author[3]{Jonathon Schwartz}
\author[2]{Michael Docking}
\author[2]{David Bowman}
\author[2]{Junae Kim}

\affil[1]{Department of Defence
\authorcr \{\tt callum.baillie\}@defence.gov.au}
\affil[2]{Defence Science and Technology
\authorcr \{\tt max.standen, david.bowman, michael.docking, junae.kim\}@dst.defence.gov.au}
\affil[3]{Australian National University
\authorcr \{\tt jonathon.schwartz\}@anu.edu.au}

\maketitle

\input{sections/introduction.tex}
\input{sections/relatedwork.tex}
\input{sections/design.tex}

\input{sections/implementation.tex}
\input{sections/evaluation.tex}

\input{sections/conclusion.tex}

\bibliographystyle{plain}
\bibliography{references}
\quad
\end{document}

%% file: sections/introduction.tex
\hyphenation{CybORG}

\begin{abstract}
Autonomous Cyber Operations (ACO) involves the consideration of blue team (defender) and red team (attacker) decision-making models in adversarial scenarios. To support the application of machine learning algorithms to solve this problem, and to encourage such practitioners to attend to problems in the ACO setting, a suitable gym (toolkit for experiments) is necessary. We introduce CybORG, a work-in-progress gym for ACO research. Driven by the need to efficiently support reinforcement learning to train adversarial decision-making models through simulation and emulation, our design differs from prior related work.
Our early evaluation provides some evidence that CybORG is appropriate for our purpose and may provide a basis for advancing ACO research towards practical applications.
\end{abstract}

\section{Introduction}

\subsection{Autonomous Cyber Operations}
Autonomous Cyber Operations (ACO) is concerned with hardening and defending computer systems and networks through autonomous decision-making and action.  We are motivated by the need to protect systems that may be isolated or operate in contested environments where skilled human operators may not be present nor have remote access, and to unlock the potential for rapid and wide scale cyber operations.

This setting has characteristic properties.  ACO is \textit{adversarial} and \textit{evolving}: behaviour is appropriate when it best achieves objectives in the face of evolving adversary behaviour (tactics, techniques and procedures, or TTPs).  Thus the setting incorporates blue team and red teams and terrain (i.e., hosts and networks).  Addressing these together requires that ACO research should be concerned with both red and blue team decision making models that may co-evolve.

Further, ACO is \textit{dynamic}, the terrain may change in uncontrolled or unpredictable ways or as the deliberate results of red and blue activity.  Importantly it is also \textit{varied}; whilst it might be interesting to solve for decision making models that are optimal for a given single scenario or configuration, computer security problems are varying and diverse.  Consider by way of metaphor that we wish to solve Chess960 (random starting positions each game) rather than standard Chess.  By generalising over diverse and dynamic scenarios, decision-making models might become resilient to variations in the environment, rendering them more adaptable and reliable.

Finally, ACO environments have high \textit{complexity} and \textit{dimensionality}, with very many interdependent and related components and large state (information relevant to decision-making and goals) and action (choices) spaces.  Given that decision-making may be considered a mapping from state to action, this is generally a harder problem for larger state and action spaces.

\subsection{Learning ACO}
These characteristics (adversarial, evolving, dynamic, varied, high complexity and dimensionality) suggest a learning approach for developing decision-making models that are trained and evaluated across a wide range of adversary behaviours and scenarios.


One possible approach is supervised learning, where a dataset provides examples of optimal decisions at each state of a game which may be used for training, validating and evaluating models.  However, due to the aforementioned characteristics, collecting relevant datasets and adapting to evolving adversaries could prove problematic for supervised learning.

An alternative approach is reinforcement learning, in which an agent \footnote{Throughout this paper we use the term agent as in \cite{RN672}, to paraphrase: a decision-making agent that seeks to achieve a goal in its environment.} learns by interacting with an environment in a trial-and-error manner guided by rewards for achieving desirable results. In this way, the algorithm learns optimal solutions by trying to maximize its reward ~\cite{RN672}. Reinforcement learning can be very effective in high complexity, high-dimensional environments.  Following this direction requires an appropriate gym or environment for training.

However, a typical problem in reinforcement learning is the requirement for a high number of samples and in many applications the resource cost and time taken to train in real world environments is prohibitive or impractical ~\cite{RN673, RN683}. Often algorithms can take considerable time and resources just to begin converging on a merely adequate solution, with an optimal one still being some distance away ~\cite{RN682}. 

The adversarial and varied setting of ACO excacerbates the problems RL has with complexity.  Training co-evolving red and blue team models and training across dynamic and varied scenarios compound these challenges.

A common strategy for overcoming this problem is to use simulations of a problem with lower fidelity and lower cost as a training ground and to then transfer the models to a higher fidelity environment \cite{RN673, RN682}. However, while simulations are often useful, they come with their own set of challenges, not the least of which is that it can be difficult to build a sufficiently accurate model to have any hope of real world application ~\cite{RN683}. Thus, the need for real world or high-fidelity, high accuracy environments remains.

Another concern when applying machine learning is overfitting. A model may produce solutions that work well for one specific problem and environment, but perform poorly when applied to similar problems or environments ~\cite{RN680, RN683, RN678, RN679}. There are numerous strategies to combat this problem including sampling multiple environments ~\cite{RN678, RN679}, and varying evaluation conditions ~\cite{RN678}, but these depend on the ability to modify and vary an environment. This further motivates the need for an ACO gym where the environments can be varied and dynamic.


\subsection{A Gym for ACO}
This work is focused on a suitable environment for intensive reinforcement learning towards ACO.
We are inspired by such toolkits such as OpenAI Gym ~\cite{1606.01540} (hence our adoption of \emph{gym}) which provide a basis for learning experiments and a set of benchmarks with which to measure progress and evaluate competing approaches.  We here propose a number of requirements appropriate for a gym for ACO research.

Firstly, we favour both simulation and emulation modes for playing through ACO scenarios.  Simulation presents the most efficient means to achieve millions of episodes for reinforcement learning necessary for agents to train, whilst emulation presents the most realistic environment within which to experience and learn such that the derived models might behave appropriately in the real world.  For intensive reinforcement learning, it may be appropriate to train a model in simulation (most efficient), and later optimise the model in emulation (most realistic), which confers an added benefit that that the simulation need not be highly accurate.  Therefore we require the ability to play the same scenario in both simulation and emulation modes, with the same APIs presented to decision-making models and learning systems.

Given that we envisage intensive reinforcement learning over millions of scenarios, it is advantageous to leverage cloud computing resources for elastic, parallel and large scale experiments.  This requires an ACO gym that may operate with and on commonly used cloud infrastructure.  Given the potential need to play millions of games to train decision-making models, the toolkit should be as efficient and low cost as possible.

Informed by the OpenAI Gym ~\cite{1606.01540} and Atari ~\cite{bellemare13arcade} environments, it is desirable for a library of scenarios or games to be available that provides a series of increasingly challenging and diverse settings and tasks as challenges and benchmarks.  Necessarily for ACO, it is an important feature that each be controllable to incorporate variability and dynamics.  We also require the scenarios to be adversarial, with red and blue teams competing to achieve their objectives and disrupt their opposition.

We envisage that such a gym would be appropriate for learning ACO decision-making models, but may also find other uses for operator training, human-machine teaming (performing with or against an ACO system), or discovering or evaluating tactics and capabilities.  Therefore it is important that the gym and its API and interfaces are easy to use and understand.

The following section presents relevant work in cyber ranges, simulation and emulation environments that have been developed for purposes including research and training operators.
The design and work-in-progress implementation of CybORG are described in Sections 3 and 4, with Section 5 providing an early evaluation considering the results of initial learning experiments.
We conclude this paper with our conclusions and plans for future work.

%% file: sections/relatedwork.tex
\section{Related Work}

\begin{table*}[t]	\centering
	\resizebox{\textwidth}{!}{ 
		\begin{tabular}{|l|l|l|l|l|l|l|l|}
			\hline
			& Simulation & Emulation & Scalable & Flexible     & Efficient & Adversarial CO & Designed for RL \\ \hline
			DETERlab~\cite{5655108}               & No         & Yes       & Low      & Yes          & Low       & No             & No              \\ \hline
			VINE~\cite{Eskridge:2015:VCE:2808475.2808486}                   & No         & Yes       & Med      & Yes          & Med       & No             & No              \\ \hline
			SmallWorld~\cite{FURFARO2018791}             & No         & Yes       & Med      & Yes          & Med       & No             & No              \\ \hline
			BRAWL~\cite{BRAWL}                  & No         & Yes       & Med      & Windows 	  & Med       & No             & Limited         \\ \hline
			Galaxy~\cite{220241}                 & No         & Yes       & Low      & Debian-based & High 	  & Limited        & Yes             \\ \hline
			Insight~\cite{Futoransky2009SimulatingCF}                & Yes        & No        & High     & Yes          & Med       & No             & No              \\ \hline
			CANDLES~\cite{Rush:2015:CAN:2739482.2768429}                & Yes        & No        & High     & Yes          & High      & Limited        & Yes             \\ \hline
			Pentesting Simulations~\cite{niculae_2018, JThesis} & Yes        & No        & High     & Yes          & Med       & Limited        & Yes             \\ \hline
			
			CyAMS~\cite{7795375}                  & Yes        & Yes       & High     & Yes          & High      & No             & No              \\ \hline
			\bf CybORG                 & \bf Yes        & \bf Yes       & \bf High     & \bf Yes          & \bf High      & \bf Yes            & \bf Yes             \\ \hline
		\end{tabular}
	}
	\caption{Existing Environments and CybORG Design vs ACO Research Requirements}
	\label{Req Table}
\end{table*}

There are a growing number of cyber security environments designed for experimentation, many of these have been considered but the requirements for both a high and low fidelity modes, the interface facilitating reinforcement learning for adversarial cyber operations and the environment being efficient flexible and scalable means that no current environment is suitable. A summary of various environments with our requirements can be found in Table \ref{Req Table}. 

DETERlab~\cite{5655108}, is a specialised cyber security experimentation environment based on EMUlab~\cite{Hibler:2008:LVE:1404014.1404023}. It supports cyber security experimentation through the emulation of hosts and networks. DETERlab has limited maximum network size because it uses a local setup of physical hardware which prevents scaling the emulation to the size of a large enterprise-like network. DETERlab is not designed to be reset quickly and can take a significant amount of time to set up a network. This is impractical for machine learning because of the number of episodes that are often required to train an agent. 

There have been several other environments inspired by DETERlab that have overcome such scalability limitations. VINE~\cite{Eskridge:2015:VCE:2808475.2808486}, SmallWorld~\cite{FURFARO2018791} and BRAWL~\cite{BRAWL} leverage scalable infrastructure such as cloud-based Infrastructure as a Service (IaaS) and efficient virtualisation frameworks such as OpenStack to emulate larger enterprise-like networks. These tools can simulate users acting on a host which are designed to generate human-like activity for the purpose of experimentation with different tools. These agents are not capable of learning and are often scripted with no ability to utilise reinforcement learning to improve their performance. BRAWL can make use of CALDERA~\cite{Applebaum:2016:IAR:2991079.2991111, CalderaPOMDP} as an red team agent to produce a more realistic experimentation testbed. It is limited to the emulation of Windows hosts only. 

GALAXY~\cite{220241} is an emulated cyber security environment which has been used to train red agents using an evolutionary algorithm to learn how to blend in reconnaissance traffic with that produced by a normal user. It has limited scalability because it currently uses a hypervisor that restricts the scenario to a single VM per physical machine. It utilises the snapshot feature to perform fast resets of VMs in an environment. 

Insight~\cite{Futoransky2009SimulatingCF} is a network simulation for experiments. It is shown to be scalable and can simulate hundreds of hosts on a single computer. Insight is only realistic from an attacker's standpoint thus only capable of facilitating red agents. This prevents the simulation from being used to perform adversarial cyber operations. Insight does not feature a reinforcement learning interface.

CANDLES~\cite{Rush:2015:CAN:2739482.2768429} leverages a high speed network security simulation to coevolve adversarial blue and red agents using an evolutionary algorithm. The red agent learns the best tactics to attack a network, however it is capable of taking unrealistic actions such as discovering an exploit which may work in the simulation but practically would take a cyber operations team a significant amount of time and resources to discover during an operation. The blue agent learns how to preconfigure the network to make it more secure, however the amount of actions blue can take is limited during the operations phase by the amount of noise made by the red agent.

Pen testing simulations~\cite{niculae_2018, JThesis} have been used to train red agents using reinforcement learning. These simulations use a realistic action space to enable agents to learn how to attack a network. The simulations can feature blue and grey agents to simulate defenders and neutral users. Blue agents act as a predefined impediment to red. Grey agents may hinder or help red by taking actions that change the state of the environment.

CyAMS~\cite{7795375} uses a combined emulation and simulation environment. CyAMS simulates a cyber security environment using a Finite State Machine (FSM). CyAMS also features an emulation and compares the simulation with the emulation using a malware propagation scenario to demonstrate the fidelity of the simulation. The aim of CyAMS is to use emulation and simulation in conjunction where a portion of the environment is emulated when higher fidelity is required otherwise simulation can be used to provide massive scalability with up to billions of clients.

CybORG aims to provide a scalable, efficient and flexible training environment that uses a high fidelity emulation and lower fidelity simulation to facilitate reinforcement learning for the coevolution of red and blue agents capable of executing cyber operations.

%% file: sections/design.tex
\section{Design}

\begin{figure*}[h]
	\includegraphics[width=1\textwidth]{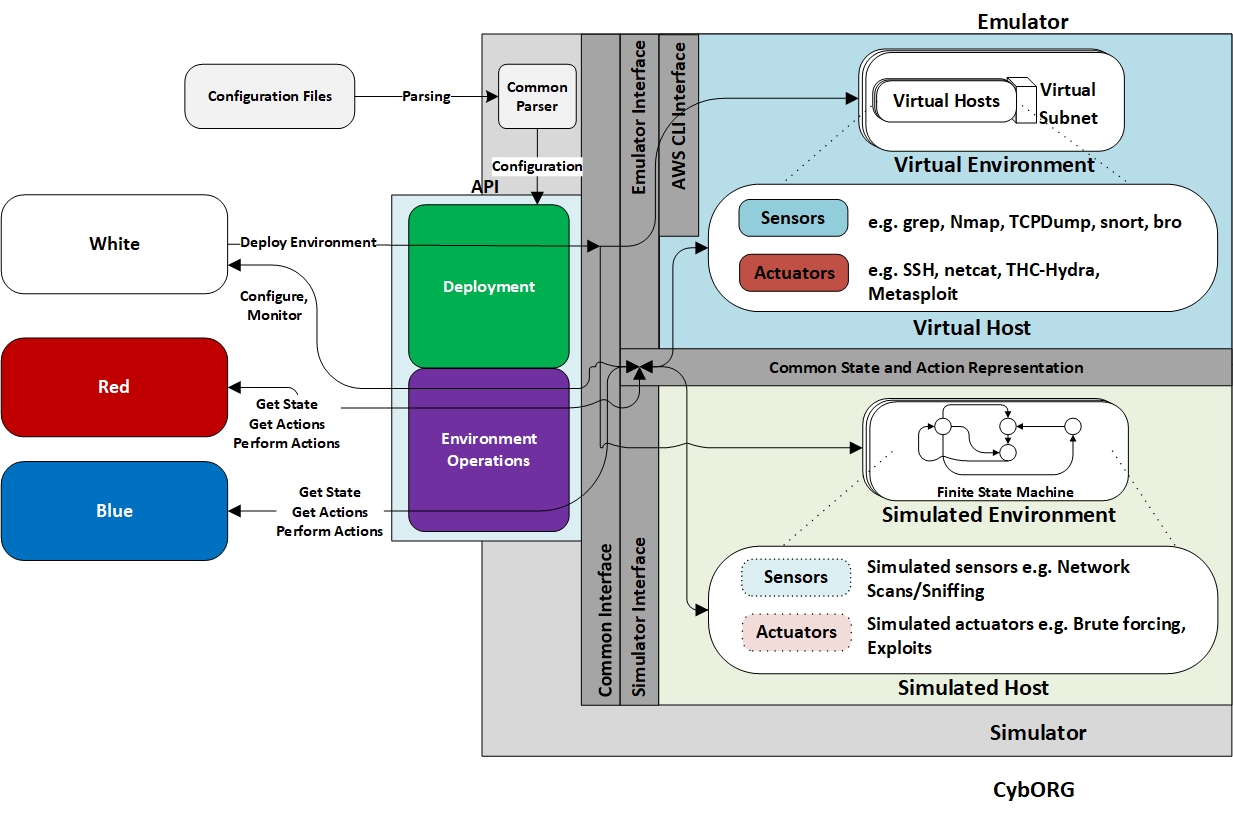}
	\centering
	\caption{CybORG Design}
	\label{fig:cyborgDesign}
\end{figure*}

This section presents the overall design for CybORG (see Figure ~\ref{fig:cyborgDesign}) incorporating simulation and emulation with a common API to support reinforcement learning for autonomous and adversarial cyber operations.

\subsection{Low-Fidelity Simulation}
CybORG's simulated environment is designed to be fast and lightweight, able to run thousands of interactions per second on a single host. This is achieved using a finite state machine design that can model networks, hosts and actions informed loosely by the real world behaviour of these components. The simulator can model the stochastic nature of some actions, but is not required to accurately capture or model real world behaviour since the emulation is available for refinement of decision-making models, maximising simulation efficiency and reducing development effort.  By keeping the simulation design lightweight, it is relatively cheap to scale.

\subsection{High-Fidelity Emulation}
To provide an accurate environment in which to perform reinforcement learning, CybORG's design incorporates a high-fidelity emulation environment. By employing Virtual Machines (VMs) deployed and organized into realistic networks, with actual cyber-related software tools in real-time, it can provide real inputs, outputs and behaviour. The emulator design allows for rapid deployment, configuration and management of numerous permutations of scenarios with varying complexity and supports cloud deployments for elastic scalability, limited only by budget.

\subsection{Common API and Interface}
CybORG's design supports reinforcement learning of a single decision-making model in both environments: a low-fidelity, efficient simulation and a high-fidelity, realistic emulation. Switching between the two environments is smoothly enabled by a common API and user interface for controlling experiments and accessing game state and action space. 

\subsubsection{Deployment}
The deployment interface provides a lightweight means of describing the desired scenario (using configuration files) that works seamlessly across both the simulation and emulation environments. In simulation, the environment is simulated entirely in software, in the emulation, the environment is deployed in AWS based on the information provided by the configuration files. By using AWS, the emulation is able to deploy multiple isolated instances of environments simultaneously.

%

\subsubsection{Cyber Operations Interface}
\begin{figure}[h]
  \includegraphics[width=0.45\textwidth]{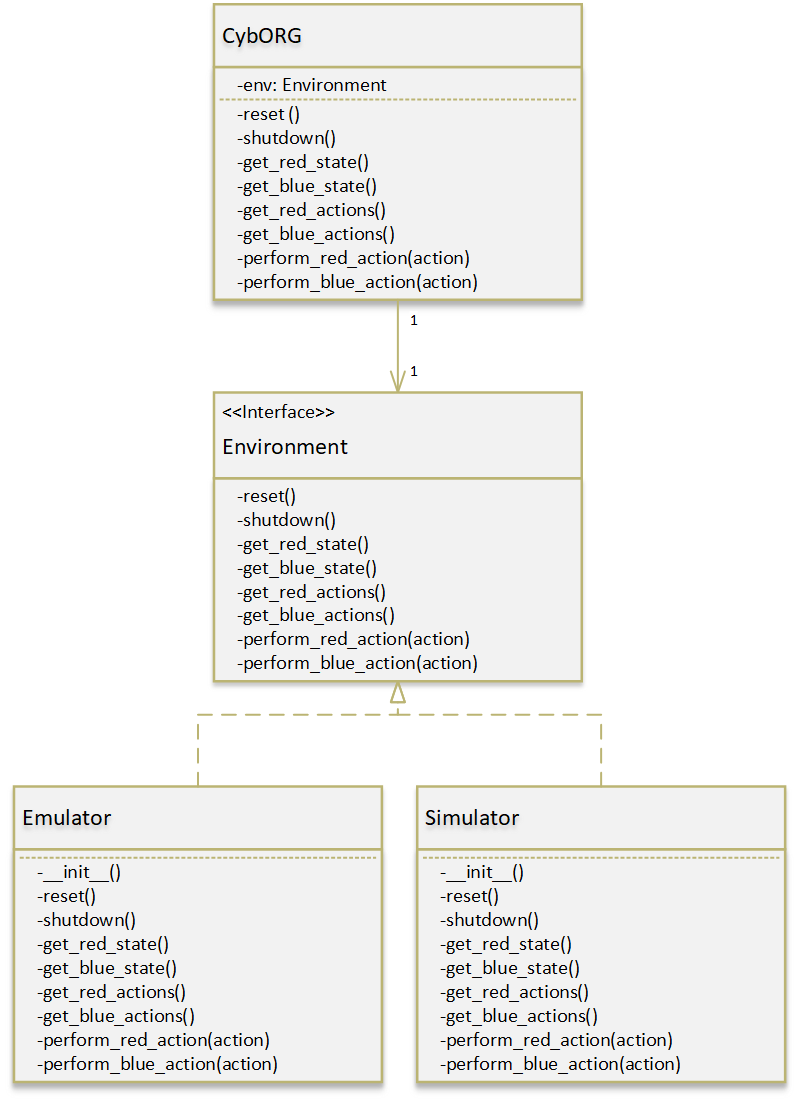}
  \centering
  \caption{The CybORG Common Interface}
  \label{fig:cyborgCI}
\end{figure}

CybORG provides an interface to enumerate the action and state space of a given environment and to execute actions. For the purposes of discovering optimal solutions, it isn't necessary for algorithms to have any notion of how an action is performed. CybORG wraps the environment and presents only the information and degrees of freedom that an algorithm needs to train. This interface is common for both the simulation and emulation environments (see Figure ~\ref{fig:cyborgCI}).

CybORG enables host interaction through the use of SSH connections to one or more hosts in the environment which act as entry points for the red, blue or white teams. From these access points, CybORG interacts with the terminal and performs actions much the same as any human operator would: by entering commands and parsing the results. This control scheme allows for multiple adversarial agents to act asynchronously and simultaneously in the environment.

%% file: sections/implementation.tex
\section{Implementation}

CybORG is in active development and new features are being implemented regularly. The details of how various components of CybORG have been implemented follow. 

\subsection{Deployment}
Deploying a particular scenario involves specifying whether it will be simulated or emulated (a flag) and providing a scenario description file (YAML). The scenario file includes details for configuring the environment including hosts, networking and subnet information, and the set of actions available to red and blue agents. The scenario files are deliberately simple, requiring a minimum of information and employing many default behaviours to reduce the burden on users when creating files.


The supported host configurations and actions are specified in separate 'Images' and 'Actions' YAML files, respectively. Parameters for hosts include "Name", "OS", "Services", "Credentials" and an ID for an image of a deployable image in AWS (for the emulation).


Action parameters describe the set of possible red and blue team actions, constituting the action space.
By combining different network architectures with selections from the available images and actions it becomes possible for users to generate and permute a wide variety of different scenarios.

\begin{figure}[h]
  \includegraphics[width=0.45\textwidth]{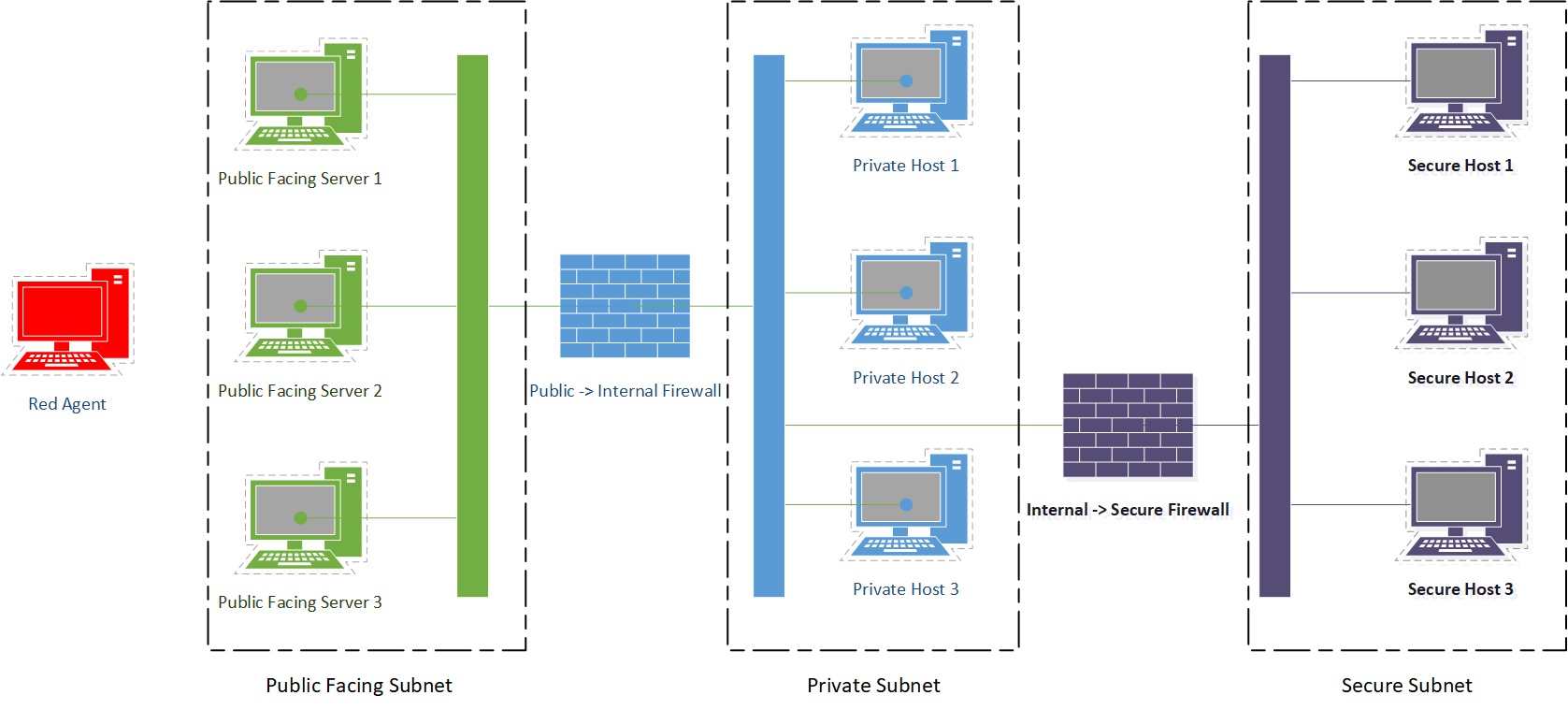}
  \centering
  \caption{A CybORG Scenario}
  \label{fig:cyborgSC}
\end{figure}

A network diagram of an example scenario that is included with CybORG and that can be deployed in both the Simulation and Emulation components of CybORG is shown in Figure ~\ref{fig:cyborgSC}.

\subsection{Emulation}
 
The emulation currently uses Amazon Web Services (AWS) with virtual machines to create a high fidelity cyber security environment, though it is intended that CybORG support a wide range of cloud/IaaS options.

\subsubsection{Deployment-Emulation}

The emulation uses the description of the scenario from the YAML scenario files to deploy and configure a virtual network in Amazon Web Services (AWS). It does this by using SSH to access a virtual gateway server in a private AWS cloud and then deploying and configuring environments using AWS's Command Line Interface (CLI) on that virtual (master) host using the following functions:

\begin{itemize}
	\item Automatically creating and deleting IPv4 subnets based on the number of IP addresses required in each subnet according to the scenario
	\item Automatically configuring routing between subnets according to the scenario
	\item Automatically creating and deleting instances of static host images and assigning them to subnets according to the scenario
\end{itemize}

Using these high-level functions, CybORG is capable of rapidly and concurrently deploying independent clusters of hosts and subnets, allowing multiple scenarios, or 'games' to be played in parallel. By using IaaS it also has the advantages those offerings provide in that it is scalable, efficient and low cost. This is especially important when conducting reinforcement learning experiments where flexible and rapid deployment of multiple parallel games is necessary to reduce overall training time.

\subsubsection{Cyber Operations}

The emulation provides an interface for red and blue agents to conduct cyber operations. CybORG implements actuators for cyber security actions and sensors for information gathering from the environment.

Though functional cyber-attack 'games' are already playable, the emulation's cyber operations functions are nascent. New actions and state information continue to be added and the depth and breadth of the cyber operations interface is improving rapidly. The environment's state properties include global properties such as known credentials, information about the network layout and subnet structure and local properties for each host including operating system information and discovered services. The environment's action space includes the ability to perform reconnaissance by using scans such as an ICMP ('ping') scan or a TCP SYN scan using Nmap or it can attack a host and obtain credentials using brute-forcing tools such as THC-Hydra. 'Blue' actions are being developed and will mirror those available in the simulation.

\subsection{Simulation}

The simulation environment of CybORG provides a lightweight and efficient model for ACO research. It is implemented as a script employing a FSM that uses classes for hosts and subnets to simply model a computing environment with security tools.

\subsubsection{Deployment-Simulation}

The simulation generates an environment from the same YAML scenario file as the emulation.
The simulation models the cyber environment as a set of subnets, hosts and actions which are abstract objects used to model the real world.
A host represents a single machine in the network and is defined by a number of attributes and actions.
Attributes of a host include its name, network address, operating system, the services it is running, credentials that it accepts and its value to an attacker.
Each subnet is defined by an address range, hosts within each subnet, routing information between itself and other subnets.

The simulator generates a single local model for each scenario instance, regardless of its size or complexity.
This has the advantage that there is no network overhead during simulation for efficiency.
The implementation is intended to maximise the number of concurrent simulations.

\subsubsection{Cyber Operations}

Simulated actions are defined by their preconditions, effects, cost and success probability.
Each action is 'executed' from a host and certain actions such as a scan may target a different object such as a subnet or another host.
The preconditions of an action are the state conditions that must be satisfied if the action is to be available.
The effects define how the action will change the state of the environment if the action is successful.
The cost is used to model the abstract cost of taking an action as part of the reinforcement learning and action selection process. This can be used to represent any metric of action performance such as time to complete or negative impact on the environment.
The success probability is used to model the sometimes stochastic nature of actions in a computing and networking environment, and defines the probability an action will succeed given all its preconditions are met.
The actions available in the simulation correspond to those available in the emulation, though the simulation currently has more actions available for the blue team. As new actions are added, they will be added concurrently to the emulation and the simulation.

The simulation maintains a hidden fully observed state of the environment and only returns the relevant information to each (red or blue) agent after each action.
In this way the simulation reflects the partial observability of the real world where a lot of information is hidden from each agent until revealed through its actions.

%% file: sections/evaluation.tex
\section{Evaluation}


In this section, we demonstrate the development and testing of reinforcement learning based-agents to perform cyber operations in the simulation.
The current state of CybORG allows for red team actions to be taken so we have used CybORG to train a red team to solve Capture the Flag (CTF) scenarios.
To accomplish this we have created CTF scenarios that feature a network of vulnerable hosts connected in different subnets.
The nature of the game is representative of a simple penetration test where a red team must exploit machines, gather intelligence and pivot through a network.




We have used Double Deep Q Networks (DDQN)~\cite{Hasselt2016Deep} to create agents and perform cyber operations on CybORG. DDQN extends Deep Q Networks (DQN)~\cite{Mnih2015} which are a deep reinforcement learning method which approximates the Q function (the expected future reward function given state and action) using deep neural networks. Unlike DQNs, DDQNs estimate Q values separately for selection and evaluation of an actions, preventing over-optimistic value estimation.

The experiment was run using a simple scenario consisting of 3 subnets with 3 hosts each as illustrated in Fig \ref{fig:cyborgSC}. One of hosts in the secure subnet has the flag which is the target of an attacking agent, a red team. The goal of the red agent is to capture the flag by obtaining privileged access to the target machine. Each machine has multiple avenues for attack, some with lower probability and some dead ends. The scenario has multiple paths/tactics leading to success for red.


A maximum reward of 1 could be obtained per episode by the agent. An episode ended when either 1000 steps had been taken or when the flag in the scenario had been captured. There were 10 independent runs with a maximum of 1000 steps per episode and 1000 episodes per run. 

\begin{figure}
	\centering
	\includegraphics[width=\linewidth]{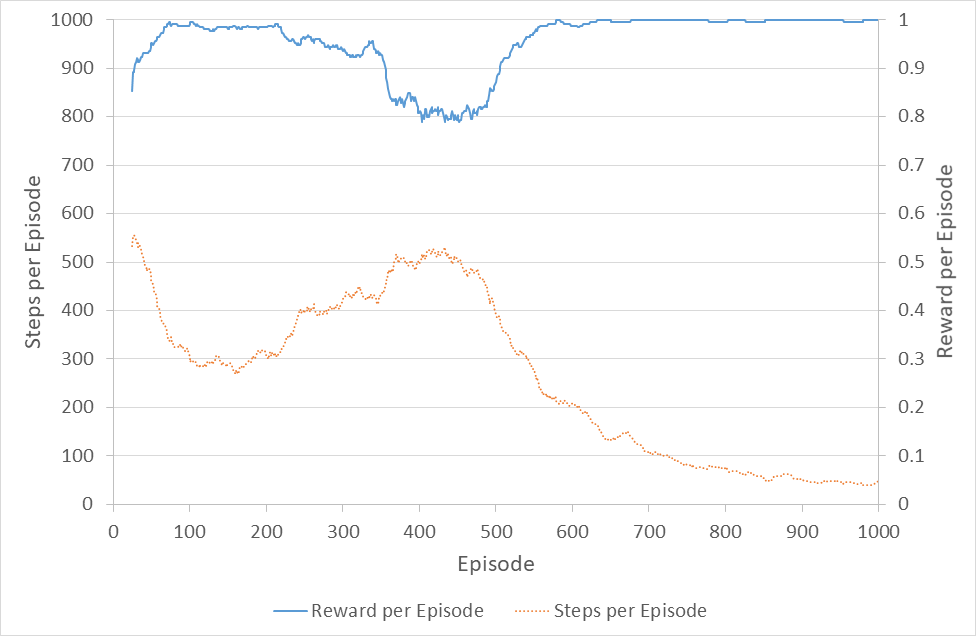}
	\caption{Results from training a DDQN agent on the scenario with 9 hosts}
	\label{RLgraph}
\end{figure}

Fig. \ref{RLgraph} illustrates the total reward and the number of steps, averaged over 25 runs, to determine the performance of the agent. The rewards are initially high due to random exploration of the state space resulting in the successful capture of the flag. The graph indicates that an agent can be trained to consistently capture all flags in the scenario.


%% file: sections/conclusion.tex
\section{Conclusion}
This paper introduced CybORG, a work-in-progress gym for Autonomous Cyber Operations research.

Whilst there have been ongoing development towards simulation and emulation environments for cyber training and/or experiments, the requirements of ACO motivate an integrated design comprising emulation and simulation modes to support large scale reinforcement learning across diverse scenarios.

We have made progress towards implementing this design, with the ability to spawn and play games either in simulation or emulation modes with cloud infrastructure, and we have performed reinforcement learning experiments with it.

CybORG or gyms like it could assist the reinforcement learning research community to apply their techniques to solve cyber operations problems, by providing an easy to use toolkit and a set of benchmark problems for guiding and measuring progress.

The primary limitation of CybORG as currently constituted is its relative immaturity.  More work is required to develop libraries of scenarios and benchmarks and to provide an easy to use interface.

\section{Future Work}
Our next step is to perform reinforcement learning with emulation mode to confirm that it functions as expected with the same interface and APIs as the simulation mode.

We then intend to perform mixed learning to train a single model under both simulation and emulation to confirm that this approach achieves efficient and effective learning with CybORG.

With just a single and simple scenario/game currently available, we intend to significantly expand the library of scenarios/games available, which once done will lead to the first public release of CybORG to the security and AI research communities.


\section*{Availability}

Limited access to CybORG will be available for reviewers, with a public release planned once the scenario library is expanded.